\documentclass[11pt,a4paper]{article}
\usepackage{jcappub}

\usepackage{amsmath}
\usepackage{amssymb}
\usepackage{tensor}
\usepackage{latexsym}
\usepackage{enumerate}
\usepackage{bbm}
\usepackage{amsthm}
\usepackage{graphicx}
\usepackage{grffile}
\usepackage{amsmath}
\usepackage{amssymb}
\usepackage{latexsym}
\usepackage{xcolor}

\usepackage[most]{tcolorbox}
\newtcolorbox{highlighted}{colback=yellow,coltext=red,breakable}
\newcommand{\ima}{\mathbbmtt{i}}
\newcommand{\vn}{\mathrm{v}}
\newcommand{\be}{\begin{equation}}
\newcommand{\ee}{\end{equation}}
\newcommand{\bea}{\begin{eqnarray}}
\newcommand{\eea}{\end{eqnarray}}
\newcommand{\pd}{\partial}
\newcommand{\mn}{{\mu\nu}}
\title{Quantum cosmology of a Bianchi III LRS geometry coupled to a source free electromagnetic field}
\author[a]{A. Karagiorgos}
\author[a]{T. Pailas}
\author[b]{N. Dimakis}
\author[a]{Petros A. Terzis}
\author[a]{T. Christodoulakis}

\affiliation[a]{Nuclear and Particle Physics Section, Physics
Department,\\
University of Athens, GR 15771 Athens, Greece}
\affiliation[b]{\it Instituto de Ciencias Fisicas y Matematicas, \\
Universidad Austral de Chile, 5090000 Valdivia, Chile}

\emailAdd{alexkarag@phys.uoa.gr}
\emailAdd{teopailas879@hotmail.com}
\emailAdd{nsdimakis@gmail.com}
\emailAdd{pterzis@phys.uoa.gr}
\emailAdd{tchris@phys.uoa.gr}

\abstract{
We consider a Bianchi type III axisymmetric geometry in the presence of an electromagnetic field. A first result at the classical level is that the symmetry of the geometry need not be applied on the electromagnetic tensor $F_{\mu\nu}$; the algebraic restrictions, implied by the Einstein field equations to the stress energy tensor $T_{\mu\nu}$, suffice to reduce the general $F_{\mu\nu}$ to the appropriate form. The classical solution thus found contains a time dependent electric and a constant magnetic charge. The solution is also reachable from the corresponding mini-superspace action, which is strikingly similar to the Reissner-Nordstr{\"o}m one. This points to a connection between the black hole geometry and the cosmological solution here found, which is the analog of the known correlation between the Schwarzschild and the Kantowski-Sachs metrics. The configuration space is drastically modified by the presence of the magnetic charge from a 3D flat to a 3D pp wave geometry.
We map the emerging linear and quadratic classical integrals of motion, to quantum observables. Along with the Wheeler-DeWitt equation these observables provide unique, up to constants, wave functions. The employment of a Bohmian interpretation of these quantum states results in deterministic  (semi-classical) geometries most of which are singularity free.}

\keywords{Quantum cosmology}

\arxivnumber{1710.02032}

\notoc
\begin{document}
\maketitle

\flushbottom

\section{Introduction}

Despite the fact that theoretical physics has been dramatically evolved in the last century, quantization of gravity still remains one of the major problems. Historically, Dirac founded the Hamiltonian formulation of General Relativity  \cite{DiracHam}, then Arnowitt, Deser and  Misner \cite{ADM} presented the canonical formulation of gravity and finally  DeWitt \cite{DeWitt} stated the well known equation.  In order to find a valid theory, several approaches have been proposed, in the following decades. The main of them are covariant(perturbation theory, path integrals) and canonical approaches (geometrodynamics \cite{kucha,isham}, connection dynamics and loop dynamics\cite{Loop1,Loop2}), and also string theory which follows a completely different path. However, in spite of the progress in each method used, there are still open issues to be addressed \cite{Kiefer}. As a result, it is more efficient to  look for certain simplification schemes, like quantum cosmology, that allow us to theoretically test each path.

The most common way to achieve the above is by using mini-superspace analysis which simulates in a simple way the quantum behaviour of certain gravitational systems with many symmetries \textbf{(for some older references see\cite{misner,hawk1,hawpag,kief,kucrya} while for more recent see \cite{San1,Bojo,Jala,Asht1,Ewing,Vak2,Bojo2,Barv,Vak1,Pal1,Pal1b,Pal2,Pal2b})}.

Thus, in this analysis the ensuing configuration is described by a finite number of degrees of freedom. Furthermore, many fundamental difficulties, arising in the quantization of full gravity (such as functional derivatives evaluated at the same spatial point and the subsequent need for regularization-renormalization), disappear. On the other hand, some of the most important properties of  full gravity, such as existence of constraints, covariance under time and space coordinate transformations are maintained \cite{chraut}; this is the reason why we can hope that mini-superspace systems could give answers for the general case.  At this level, the major paths that can be followed, are two: the standard canonical quantum mechanics or the polymer quantization \cite{Corichi,Date}, which has been put in use in the framework of Loop Quantum Cosmology.

In the present work the standard canonical quantization is employed. In  this scheme  the classical symmetries of constrained systems \cite{tchrisJP} are  promoted to operators and used, along with the Wheeler-DeWitt equation, as extra conditions on the wave functions \cite{tchrsSch}. In this way quantum observables and their eigenvalues are correlated to classical constants of integration which appear in the metric. This method has been used in a variety of  cosmological configurations \cite{tchriscosmo} as well as black holes \cite{tchrsSch,tchrisRN,tchrisBTZ}. A noteworthy fact of this method is that in some specific subalgebras, the quantization  leads to semiclassical avoidance of curvature singularities, which is one of the main expectations in quantum cosmology. What is more, one can invoke the parametrization invariance of such systems to generalize the quantization procedure upon a certain class of geometries \cite{apato}.

Our analysis is focused on the quantization with respect to Abelian subalgebras of the symmetries, which is a quantization procedure initially presented in \cite{tchrisRN}. The main difference here is that we also include possibly existing higher order symmetries (like second rank non-trivial Killing tensor fields) in the search for the desirable Abelian subalgebras.  The representation of these higher symmetries at the quantum level is made through a pseudo-Laplacian operator \cite{Carter,benenti2002,Duval}.

In this work, we study the classical dynamics and the canonical quantization of a model consisting of a Bianchi type III LRS  geometry  coupled to a source-free electromagnetic field. Previous works concerning Bianchi III plus some fields or perfect fluids include \cite{mimoso,carn,koiv}. However, the matter content is scalar fields and/or perfect fluids, with the general assumption of being shear free which translates into a FLRW like scale factor matrix $\gamma_{\alpha\beta}=diag(a(t),a(t),a(t))$. Thus, even if the matter content could be parameterize as that of an electromagnetic field,  the solution would only be a subject of the solution space here found. Furthermore, none of them is treated quantum mechanically. A very important result is that the emergent mini-superspace configuration space is drastically modified by the presence of the magnetic field from a flat 3D geometry (electric field only) to a pp-wave 3D geometry (electric plus magnetic field). As a consequence of this all Type I curvature scalars(e.g. $g^{\mu\nu}R_{\mu\nu}, R^{\mu\nu}R_{\mu\nu}, R_{\mu\nu\kappa\lambda;\sigma}R^{\mu\nu\kappa\lambda;\sigma}$,...) vanish, while second order non-trivial symmetries appear. The use of the existing conserved quantities enables us to easily derive the classical solution which, interestingly enough, produces the same  geometrical family as in the case of a single electric field with $Q^2+\epsilon^2$ replacing the charge $Q^2$ appearing in the case where only electric field is considered (see also the black hole case \cite{dimet2017}).

The structure of the paper is as follows: at first a mini-superspace reduction and a canonical quantization scheme  are presented. In the third section we study the classical dynamics and the canonical quantization on the three dimensional mini-superspace metric that arises from the Bianchi type III LRS space-times in the presence of an electric and a magnetic field. The cases of the single magnetic field and/or electric field are also treated in sections 4 and 5 respectively. In each of these cases  a semi-classical analysis of the solutions is done using Bohm' s interpretation of quantum mechanics. In this limit we find, at some cases  singularity free space-times. Lastly, a discussion of the results is given.

\section{Mini-superspace models and  canonical quantization process} \label{section1}

Our strating point is the action of Einstein's gravity plus matter
\begin{equation} \label{generalact}
  S= \frac{c^4}{16\pi G} \int\!\! \sqrt{-g}\, R\, d^4 x + S_m ,
\end{equation}
where $g$ is the determinant of the space-time metric $g_{\mu\nu}$, $R$ the Ricci scalar and $S_m$ the action of the matter content. In the case that  the manifold has a specific group of isometries, such as  spatial homogeneity, the line element can be decomposed in the following form
\begin{equation}\label{generalline}
  ds^2 = -N(t)^2 d t^2 + \gamma_{\kappa\lambda}(t) \sigma^{\kappa}_i(x) \sigma^{\lambda}_j(x) dx^{i} dx^{j},
\end{equation}
where $N$ is the lapse function. Note that the more general line element should also have a shift term $(2N_\alpha\sigma^{\alpha}_i(x)dx^i dt)$ which can always be made to vanish through a particular coordinate transformation \cite{chraut} (however, one has to keep in mind that it severel cases it may so happen that this absorbtion can be performed only locally, since it can results in altering the topology of the
time direction.). If one inserts this line element into the Einstein's equations (obtained by varying the action with respect to $g_{\mu\nu}$)
\begin{equation}\label{generalEin}
  G_{\mu\nu}\equiv R_{\mu\nu} - \frac{1}{2} g_{\mu\nu} R = \frac{8\pi G}{c^4} T_{\mu\nu} ,
\end{equation}
with $T_{\mu\nu} = \frac{2}{\sqrt{-g}} \frac{\delta S_m}{\delta g^{\mu\nu}}$, one arrives at  a set of coupled ordinary differential equations with $t$ as the independent dynamical variable. On the other hand, if the action  \eqref{generalact} is calculated using the decomposed metric \eqref{generalline}, integrating out the non-dynamical degrees of freedom, the full gravitational system may be successfully described  by the reduced action. If this happens this action is called valid. This  reduced system has finite degrees of freedom and it is thus more easily quantized.

The Lagrangian\footnote{For simplicity in what follows, we choose to work in units $c=G=\hbar=1$, $\mu_0=4 \pi$. The choice for $c$, $G$ and $\hbar$ reflect the typical geometric and quantum mechanical units respectively, while the choice for $\mu_0$ means that we are working in Gaussian units for the electromagnetic field.} of these mini-superspace systems has the general form
\begin{equation}\label{generalLag}
  L= \frac{1}{2 N(t)} \overline{G}_{\alpha\beta}(q) q'^{\alpha}(t) q'^{\beta}(t) - N(t) V(q),
\end{equation}
where $' = \frac{d}{dt}$, while $V(q)$ and $\overline{G}_{\alpha\beta}(q)$ are the mini-superspace potential and metric respectively. The variables $q^{\alpha}$ are functions of the matrix scale factors $\gamma_{\mu\nu}$ i.e. $q^{\alpha}=q^{\alpha}(\gamma_{\mu\nu})$ . This Lagrangian is corresponding to a singular system. The associate Hamiltonian is written as
\begin{equation*}
  H_T = N \mathcal{H} + u_N p_N ,
\end{equation*}
which is produced with the help of the Dirac-Bergmann algorithm for singular systems \cite{Dirac,AndBer} . This Hamiltonian has the following primary and secondary constraints
\begin{subequations}
\begin{align}
  p_N & \approx 0, \\ \label{Hamcon}
  \mathcal{H} =\frac{1}{2} \overline{G}^{\alpha\beta} p_\alpha p_\beta + V(q)  & \approx 0.
\end{align}
\end{subequations}

In what follows, we will use the constant potential parametrization in which all the information about the system is included inside the mini-superspace metric. This parametrization is achieved if we adopt the following transformation $N \mapsto n = N\, V$, thus the transformed Lagrangian is
\begin{equation}\label{generalLag2}
  L= \frac{1}{2 n(t)} G_{\alpha\beta}(q) q'^{\alpha}(t) q'^{\beta}(t) - n(t),
\end{equation}
with $G_{\alpha\beta} = V\, \overline{G}_{\alpha\beta}$ being the new mini-superspace metric. The corresponding Hamiltonian constraint becomes
\begin{equation*}
  \mathcal{H} =\frac{1}{2} G^{\alpha\beta} p_\alpha p_\beta + 1   \approx 0 .
\end{equation*}

It is easily proven that there can be conserved quantities (linear in momenta), modulo the constraint \cite{Kuchar}, with the general form
\begin{equation}\label{conpotcons}
  Q= \xi^{\alpha}(q) p_\alpha + \int\!\! n(t) \omega(q(t)) dt ,
\end{equation}
with
\begin{equation*}
  \mathcal{L}_\xi G_{\alpha\beta} = \omega(q) G_{\alpha\beta} ,
\end{equation*}
where $\mathcal{L}_\xi$ is  the Lie derivative with respect to the configuration space vector $\xi$.
The above conserved charges  correspond to conformal ($\omega(q)\neq 0$), Killing ($\omega(q)=0$) or homothetic ($\omega(q)=\text{constant}$) vector fields of $G_{\alpha\beta}$. The charges which correspond to Killing veftor fields  strongly commute with the Hamiltonian, a property which will be extremely useful in the process of quantization, as we shall show in the following analysis.

In the system under consideration there could arise  non-trivial higher order symmetries. These symmetries are generated by second order Killing tensors, defined through
\be\label{killfield}
\nabla_\mu K_{
\nu \lambda }+\nabla_\lambda K_{\mu
\nu }+\nabla_
\nu K_{\lambda \mu}=0 ,
\ee
\be
K_{\mu\nu}=K_{\nu\mu} .
\ee
These tensors are separated in two categories: the first includes the  trivial tensors constructed by tensor products of Killing vector fields
\be
K_{\mu\nu}=\frac{1}{2} \left( \xi_\mu\otimes\xi_\nu+\xi_\nu\otimes\xi_\mu \right)
\ee
or the metric itself, while the second contains all the rest, which are designated as non-trivial.

The corresponding phase space quantities $K=K^{\mu\nu}p_\mu p_\nu$ have vanishing Poisson brackets with the Hamiltonian constraint, and therefore constitute constants of motion
\be\label{qtcon}
K=K^{\mu\nu}p_\mu p_\nu\Rightarrow
\{K,\mathcal{H}\}=0 .
\ee

The next step is to construct a canonical quantization scheme for the Lagrangian \eqref{generalLag2}. We assume that the mini-supermetric $G_{\alpha\beta}$ possesses some Killing vector fields $\xi_I$ and some Killing tensor fields $K_{J}$ where $I, \ J$ are  indices which label each one of them. To each Killing vector  and  tensor field corresponds an integral of motion. We begin by assigning differential operators to momenta and replace Poisson brackets by commutators
\begin{equation*}
  p_n \mapsto \widehat{p}_n = -\ima\, \frac{\partial}{\partial n}, \quad p_\alpha \mapsto \widehat{p}_\alpha = - \ima \, \frac{\partial}{\partial q^\alpha},\quad \{ \ , \  \}\rightarrow-\ima  [  \ , \  ] ,
\end{equation*}
while the operators corresponding to $q^\alpha$ are considered to act multiplicatively. In order to solve the factor ordering problem of the kinetic term of $\mathcal{H}$, we choose the conformal Laplacian (or Yamabe operator),
\begin{equation}\label{Hamoperator}
  \widehat{\mathcal{H}} = -\frac{1}{2 \mu} \partial_\alpha \left(\mu G^{\alpha\beta}\partial_\beta \right)+ \frac{d-2}{8(d-1)} \mathcal{R} + 1,
\end{equation}
where $\mu(q)= \sqrt{|\det{G_{\alpha\beta}}|}$, $\partial_\alpha = \frac{\partial}{\partial q^\alpha}$, $\mathcal{R}$ is the Ricci scalar and $d$ the dimension of the mini-superspace. This choice is essentially unique on account of:
\begin{enumerate}
\item
The requirement for the operator to be scalar under coordinate transformations of the configuration space variables $q^a$.
\item
The requirement to contain second derivatives of $G_{\mu\nu}$ since the classical constraint is quadratic in momenta.
\item
The requirement to be covariant under conformal scalings of $G_{\mu\nu}$, since this is also a property of the classical system.
\end{enumerate}
A further property of the definition \eqref{Hamoperator} is that it will be Hermitian in the Hilbert  space of the quantum states, if of course appropriate boundary conditions are fulfilled e.g. square integrability of the derivatives of $\Psi(q)$ \cite{vilen}. However, in view of the fact that any particular combination of the $q^a$'s can, at the classical level, be considered as representing the time, such a property is not in general expected to hold. This can be considered as a reflection of the famous problem of time in quantum gravity/cosmology \cite{isham,kucha}.

As far as the classical symmetries (\ref{conpotcons} and \ref{qtcon}) are concerned, they can naturally be transfered to operators by  formally assigning to $Q_I$ the general expression for linear first order, Hermitian operators \cite{Dim1} and to $K_J$ a pseudo-Laplacian operator \cite{benenti2002}; thus the corresponding forms are respectively
\begin{equation}\label{firstordop}
  \widehat{Q}_I = -\frac{\ima}{2\mu} \left( \mu \xi_I^{\alpha} \partial_\alpha + \partial_\alpha (\mu \xi_I^{\alpha})\right) = -\ima\, \xi_I^\alpha \partial_\alpha ,
\end{equation}
\be \label{Kop}
\widehat{K}_J=-\frac{1}{\mu}\partial_\alpha\left[\mu K_J^{\alpha\beta}\partial_\beta\right].
\ee
In \eqref{firstordop} the last equality holds due to the $\xi_I$'s being Killing vector fields. The linear symmetries exactly commute with the Hamiltonian  only in the constant potential parametrization \cite{tchrsSch} i.e. commutators  of the following form are zero
\begin{equation} \label{comQH}
  [\widehat{Q}_I,\widehat{\mathcal{H}}] = 0 .
\end{equation}

By virtue of the above relations, it is possible to use some of the $\widehat{Q}_I$'s as quantum observables together with $\widehat{\mathcal{H}}$ as long as they have a common set of eigenfunctions.
Therefore, our quantum  system will obey the following conditions
\begin{subequations} \label{quantcon}
\begin{align}
  \widehat{p}_n\Psi & = 0 \\ \label{WDW}
  \widehat{\mathcal{H}} \Psi & =0\\
\label{eigeneq}
  \widehat{Q}_I \Psi & = \kappa_I \Psi
\end{align}
\end{subequations}
where the first is restriction induced by the primary constraint and the second one is the Wheeler-DeWitt equation.
The number of $\widehat{Q}_I$ that can be consistenly imposed on the wave function is  prescribed by the integrability condition \cite{tchrisBI}
\begin{equation*}
  C^M_{IJ} \kappa_M = 0 ,
\end{equation*}
where $\kappa_M$ are the eigenvalues and $C^M_{IJ}$ the structure constants of the subalgebra under consideration.

Similar considerations apply when some of the $\widehat{K}_J$ are also used. In order for these quadratic quantum observables to be consistently imposed there are geometric conditions involving the metric and the $K_J^{\alpha\beta}$' s that need to be satisfied \cite{benenti2002}.

\section{The general case with electric and magnetic field }

In this section we study the canonical quantization of the  homogeneous LRS Bianchi type III space-time in the presence of an electromagnetic field.  As mentioned in the introduction  the emergent configuration space is a pp-wave, i.e. contains one covariantly constant, null vector field. Initially, we construct the classical system and its valid reduced action.  Afterwards we proceed to the quantization of this system and finally we give  solutions corresponding to the quantum algebra of three conserved quantities which are the quadratic quantum constraint and the quantum analogues of a linear and a quadratic conserved charge.

\subsection{Classical Description}
At the beginning we consider the action \eqref{generalact} with the matter content
\begin{equation*}
  S_m = -\frac{1}{16 \pi}\int\!\! \sqrt{-g}  F^{\mu\nu}F_{\mu\nu} d^4 x .
\end{equation*}
 As it is well known the variation of the above action with respect to the metric, yields the following energy - momentum tensor

\begin{equation*}
  T_{\mu\nu} =  \frac{1}{4 \pi}\left(F_{\mu\kappa}F_{\nu}^{\phantom{\nu}\kappa}-\frac{1}{4} F^{\kappa\lambda}F_{\kappa\lambda} g_{\mu\nu}\right) ,
\end{equation*}
where $F_{\mu\nu}$ is  the electromagnetic tensor, defined in terms of the electric and magnetic fields as $F_{\mu\nu}=-F_{\nu\mu}$, $F_{0i}=E_i(t,x,y,z)$ and $F_{ij}=\epsilon_{ijk}B^k(t,x,y,z)$. Furthermore, by virtue of the equivalence principle, this tensor must satisfy the  (source free)  Maxwell equations
\begin{equation}\label{Maxwell1}
  F^{\mu\nu}_{\phantom{\mu\nu};\nu} = 0 ,
\end{equation}
\be\label{Maxwell2}
\epsilon^{\mu\nu\lambda\kappa}F_{\kappa\lambda;\nu}=0 .
\ee

In addition,  the generic line element for a Bianchi Type III LRS \footnote{A space-time is called locally rotationally symmetric (LRS) if at each point $p$ in an open neighborhood $U$ of a point $p_o$, there exists a nondiscrete subgroup $g$ of the Lorentz group in the tangent space $T_p$ which leaves invariant the curvature tensor and all its covariant derivatives to the second order, i.e. \cite{Ellis1967,EllisMacCallum1969,Goode1986}} model is
\begin{equation} \label{twometr}
    ds^2 =  -N(t)^2 dt^2 + a(t)^2 \left(dx^2 + e^{-2 x} dy^2 \right) + b(t)^2 dz^2
\end{equation}
and admits the following four Killing fields
\be\label{killfield}
\xi_1=\pd_y \qquad \xi_2=\pd_z \qquad\xi_3=\pd_x+y \pd_y\qquad\xi_4= 2y \ \pd_x+(y^2-e^{2x})\pd_y ,
\ee
with an emerging Einstein tensor satisfying the following algebraic conditions
\be
G_{\mu\nu}=0 \quad\text{for} \quad\mu\neq\nu \qquad \text{and} \qquad
G_{11}=G_{22}e^{2x} .
\ee
The Einstein equations \eqref{generalEin} dictate that the same algebraic relations must also be satisfied by the energy-momentum tensor $T_\mn$ given above. The system is quantratic in $E_i$, $B_j$ and admits  five different solutions, with only one giving real valued  electromagnetic field $F_\mn$.
This solution is described by $E_1=E_2=B_1=B_2=0$,
thus resulting in the following electromagnetic tensor
\be\label{emtensorin}
F_{\mu\nu}=\left(
\begin{array}{cccc}
 0 & 0 & 0 & E_3(t,x,y,z) \\
 0 & 0 & B_3(t,x,y,z) & 0 \\
 0 & -B_3(t,x,y,z) & 0 & 0 \\
 -E_3(t,x,y,z) & 0 & 0 & 0 \\
\end{array}
\right) .
\ee
At this stage repeated and combined use of Maxwell' s equations (\ref{Maxwell1},\ref{Maxwell2}) results  into the final form of electromagnetic  tensor which is described by setting $E_3(t,x,y,z)=E_3(t)$ and $B_3(t,x,y,z)=\epsilon \ e^{-x}$, with $\epsilon$ being a constant of integration. The only remaining Maxwell equation is
\be
F^{3\nu}_{\phantom{\mu\nu};\nu} =\frac{2 a'(t)}{a(t)}-\frac{b'(t)}{b(t)}+\frac{E_3'(t)}{E_3(t)}-\frac{N'(t)}{N(t)}=0
 \ee
and by solving it  for $E_3(t)$ gives
\be
E_3(t)=Q\frac{ b(t) N(t)}{a(t)^2} ,
\ee
where $Q$ is an integration constant. Subsequently, if we substitute this result into \eqref{generalEin}, integrate equation $G_{33}=T_{33}$ to obtain $N(t)$, insert this result into the rest of Einstein' s equations and choosing the gauge to be $a(t)=t$, we arrive to the following simple equation
\be
b'(t) \left(\epsilon^2+ \left(2 m-t\right)t+Q^2\right) t + b(t) \left(\epsilon^2 + m t+Q^2\right)=0,
\ee
where $m$ is a constant of integration. The above can be readily solved with the final line element and $F_\mn$ reading (after discarding constants of integration that are not essential for the system)
\be\label{lineemel}
ds^2=- \left(1- \frac{2 m}{t} - \frac{Q^2+\epsilon^2}{t^2} \right)^{-1}dt^2+
 t^2 \left(dx^2+
e^{-2 x}  dy^2 \right)
+\left(1- \frac{2 m}{t} - \frac{Q^2+\epsilon^2}{t^2} \right)dz^2 ,
\ee
\be\label{eltensorred}
F_\mn=\left(
\begin{array}{cccc}
 0 & 0 & 0 & \frac{Q}{t^2} \\
 0 & 0 & \epsilon \ e^{-x} & 0 \\
 0 & -\epsilon \ e^{-x} & 0 & 0 \\
 -\frac{Q}{t^2} & 0 & 0 & 0 \\
\end{array}
\right) .
\ee
To the best of our knowledge this line element has not been previously appear in the literature.  There are the following invariant relations characterizing this geometry:
\begin{subequations}
    \begin{align}
      R&=0 \\
     W_{;\kappa}W^{;\kappa}&=\frac{64 \sqrt[4]{2} m W^{19/8}}{Qe^{3/2}}-\frac{32 \sqrt{2} W^{9/4}}{Qe}+32 W^{5/2}  \\
   \Box W &=\frac{64 \sqrt[4]{2} m W^{11/8}}{Qe^{3/2}}-\frac{28 \sqrt{2} W^{5/4}}{Qe}+36 W^{3/2}\\
R_{\mu\nu\kappa\lambda}R^{\mu\nu\kappa\lambda}&=\frac{12 \sqrt{2} m^2 W^{3/4}}{Qe^3}+\frac{24 \sqrt[4]{2} m W^{7/8}}{Qe^{3/2}}+14 W
    \end{align}
  \end{subequations}
where $W=R_{\mu\nu}R^{\mu\nu}$  is the square of the Ricci tensor. The first relation is common to all solutions of the Einstein-Maxwell system, since the trace of the corresponding energy momentum tensor is zero in four dimensions. The other relations can be used to verify that the previously found solutions in \cite{mimoso,carn,koiv} are indeed different geometries than the one described by \eqref{lineemel}. This also applies for the cylindrically symmetric colliding wave solutions of Chandrasekhar and Xanthopoulos \cite{chandraxant,chandra}.

From the line element \eqref{lineemel} we can observe the analogies that exist to the Reissner–Nordstr\"om geometry, as also happens between the Schwarzschild and the vacuum Kantowski-Sachs solution. The spacetime exhibits a curvature singularity at $t=0$ and a coordinate singularity at
\begin{equation}
  t= t_{\pm} = m \pm \sqrt{m^2 + Q^2 +\epsilon^2}.
\end{equation}
In the region $t \in \mathbb{R}^{+}$ there always exists a horizon, instead of two, one or none of the Reissner–Nordstr\"om case, where we have $r \in \mathbb{R}^{+}$ for the radial distance. This is owed to the different sign with which there appear the squares of the two charges in \eqref{lineemel}. We should also note that the constant of integration $m$ is not necessarily related to a mass of some object, so we may consider it either positive or negative.

We next wish to obtain a mini-superspace description of the above system. The usual way to do this is to apply the symmetries of the geometry to the matter fields which in our case are incorporated in $F_\mn(t,x,y,z)$. The demand that the Lie derivative vanish for all  four Killing fields \eqref{killfield} i.e. $\mathcal{L}_{\xi_i}F_{\mn}=0$, results in the general form \eqref{emtensorin} with the specializations $E_3(t,x,y,z)=E_3(t)$ and $B_3(t,x,y,z)=\epsilon(t) \ e^{-x}$. At this stage, we consider the fact that $F_\mn$ should be given in terms of a potential 1-form $A_\mu$ i.e. $F_{\mu\nu} = \partial_\mu A_\nu - \partial_\nu A_\mu$. This implies that $\epsilon(t)$ must be a constant and the corresponding 1-form reads $A = -\epsilon \,e^{-\mathrm{x}} dy+f(t)dz$.
It is noteworthy that if we had kept the magnetic field $\epsilon(t)e^{-x}$ , the resulting mini-superspace Lagrangian would be  invalid i.e. its equations of motion would differ from  Einstein's plus Maxwell.

In a manner similar to the case that we analyzed in the introduction, we produce the total valid mini-superspace Lagrangian, which reads
\begin{equation}
L=
\frac{-2 a(t) b(t) a'(t) b'(t)- b(t)^2 a'(t)^2+a(t)^2 f'(t)^2}{b(t) N(t)}-N(t)\frac{b(t) \left(a(t)^2+\epsilon ^2\right)}{a(t)^2} .
\end{equation}
As previously discussed we go over to the constant potential parametrization with the change
\begin{equation} \label{ntoN2}
  N(t) \mapsto n(t) = N(t)\frac{b(t) \left(a(t)^2+\epsilon ^2\right)}{a(t)^2} .
\end{equation}
Thus, the Lagrangian is transformed into the following form
\begin{equation}\label{lagconpot}
  L =\frac{\left(a(t)^2+\epsilon ^2\right) \left(-2 a(t) b(t) a'(t) b'(t)-b(t)^2 a'(t)^2+a(t)^2 f'(t)^2\right)}{a(t)^2 n(t)}-n(t).
\end{equation}
The corresponding mini-superspace metric is
\begin{equation}\label{metricab}
  G_{\mu\nu}= \frac{2b}{a} \left(a^2+\epsilon ^2\right)\left(
\begin{array}{ccc}
 -\frac{b}{ a} & -1 & 0 \\
 -1 & 0 & 0 \\
 0 & 0 & \frac{a}{b} \\
\end{array}
\right) ,
\end{equation}
which has a vanishing Cotton-York tensor, i.e it is a conformally flat geometry. Moreover, the metric admits three Killing vector fields and one homothetic
\be \label{symvecpp}
\xi_1=\partial_f, \qquad \xi_2=\frac{1}{ab}\partial_b, \qquad \xi_3=\frac{f}{ab}\partial_b-\frac{1}{a}\partial_f, \qquad
\xi_h=\frac{b}{2}\partial_b+\frac{f}{2}\partial_f
\ee
of which $\xi_2$ is null and has vanishing covariant derivative, revealing the nature of the geometry as a pp-wave.

It is straightforward to derive the classical solution by using the conserved quantities of the form $Q_I = \xi_I^\alpha p_\alpha$ expressed with the help of the three Killing vector fields plus the one stemming from the homothetic vector field which reads
\begin{equation}
  Q_h = \xi_h^\alpha p_\alpha + \int\!\! n(t) dt.
\end{equation}
The general solution can be given in terms of $n(t)$, $f(t)$ and $b(t)$
\begin{equation}
\begin{split}\label{solut}
&  n(t) = -\frac{2 a'(t)(a(t)^2+ \epsilon^2)}{k_2 a(t)^2}, \quad f(t) =\frac{k_1}{k_2a(t)} + \frac{k_3}{k_2}, \\
& \quad b(t) = \pm \frac{\sqrt{ k^2_2 n_1 a(t)+4 a(t)^2- (k_1^2+ 4 \epsilon ^2)}}{k_2 a(t)} ,
\end{split}
\end{equation}
where $k_1, \ k_2, \ k_3$ and $n_1$ are integration constants, with the $k_i$'s being those that correspond to the conserved charges $Q_i$. If we would like to acquire the solution \eqref{lineemel} when $a(t)=t$ we have to choose $k_1=2 Q$, $k_2=2 c_2$, $n_1=-2 m/c_2$ and absorb $c_2$ with a scaling in the $z$ variable. The function $a(t)$ remains arbitrary since a gauge fixing choice need not be applied in order to solve the system of equations at hand. By using relation \eqref{ntoN2} and substituting it to \eqref{twometr} we are naturally led to the following line element
\be\label{linelcl}
\begin{split} ds^2 = - \left(1- \frac{2 m}{a(t)} -\frac{Q^2 +\epsilon^2}{a(t)^2}\right)^{-1} a'(t)^2 dt^2 & +  a(t)^2 \left(dx^2 +  e^{-2 x} dy^2 \right)  \\
& + \left(1- \frac{2 m}{a(t)} -\frac{Q^2 +\epsilon^2}{a(t)^2}\right) dz^2 .
\end{split}
\ee
In \cite{dimet2017} the case of the Reissner-Nordstr\"om geometry has been considered. The classical solution found there is obtained by a Lagrangian strikingly similar to \eqref{lagconpot} with $r$ as the dynamical variable. The  difference with the present case is  that now  the metric contains the combination $Q^2+\epsilon^2$ instead of $Q^2$, thus the geometry has two essential constants, $Q^2+\epsilon^2$ together with $m$. At this point we can note that, although we derive the same solution classically, the fact that the geometry of the mini-superspace has changed leads to a different quantization procedure. This is due to the fact that the quantum theory can distinguish between two different matter configurations despite the fact that they lead to the same base manifold geometry.

Before we proceed with the quantization it is useful to go over to the variables that bring into normal form two of the commuting Killing vector fields $\xi_1$ and $\xi_2$ of \eqref{symvecpp}. Hence, we apply the transformation $(a,b,f)\mapsto(\chi,\psi,\zeta)$ with
\be \label{tranppw}
a = \chi, \qquad b=\sqrt{2\frac{\psi}{\chi}}\qquad f= \zeta
\ee
and the Killing vector fields expressed in the new variables are
\be \label{kilplushomxyz}
\xi_1=\partial_\zeta\qquad\xi_2=\partial_\psi\qquad \xi_3=\zeta\partial_\psi-\frac{1}{\chi}\partial_\zeta \qquad\xi_h=\psi\partial_\psi+\frac{\zeta}{2}\partial_\zeta ,
\ee
while the corresponding super-metric is
\be
G_{\mu\nu}=2(x^2+\epsilon^2)\left(
\begin{array}{ccc}
 0 & -\frac{1}{x^2} & 0 \\
-\frac{1}{x^2}  & 0 & 0 \\
 0 & 0 & 1 \\
\end{array}
\right) .
\ee

Additionally to \eqref{kilplushomxyz} there also exist non-trivial higher order symmetries. In particular, there are five irreducible second rank Killing tensor fields (i.e. satisfying the relation \eqref{killfield}) which, in $(\chi,\psi,\zeta)$ coordinates have the following matrix form
\be
\begin{split}
 K_1^{\mu\nu}=\left(
\begin{array}{ccc}
 \frac{\chi^2}{2} & \frac{\chi \psi \left(\epsilon ^2-\chi^2\right)}{2 \left(\chi^2+\epsilon ^2\right)} & 0 \\
 \frac{\chi \psi \left(\epsilon ^2-\chi^2\right)}{2 \left(\chi^2+\epsilon ^2\right)} & \frac{\psi^2}{2} & 0 \\
 0 & 0 & \frac{\chi \psi}{\chi^2+\epsilon ^2} \\
\end{array}
\right), & \\
K_2^{\mu\nu}=\left(
\begin{array}{ccc}
 0 & \frac{\chi \zeta \left(\chi^2-\epsilon ^2\right)}{2\left(\chi^2+\epsilon ^2\right)} & \frac{1}{2} \\
 \frac{\chi \zeta \left(\chi^2-\epsilon ^2\right)}{2\left(\chi^2+\epsilon ^2\right)} & - \psi \zeta & \frac{ \psi}{2\chi} \\
 \frac{1}{2} & \frac{\psi}{2 \chi} & -\frac{\chi \zeta}{\chi^2+\epsilon ^2} \\
\end{array}
\right), & \nonumber
\end{split}
\ee

\be
K_3^{\mu\nu}=\left(
\begin{array}{ccc}
 0 & \frac{\chi \left(\epsilon ^2- \chi^2\right)}{2 \left(\chi^2+\epsilon ^2\right)} & 0 \\
 \frac{\chi \left(\epsilon ^2- \chi^2\right)}{2 \left(\chi^2+\epsilon ^2\right)} & \psi & 0 \\
 0 & 0 & \frac{\chi}{\chi^2+\epsilon ^2} \\
\end{array}
\right),
\quad
K_4^{\mu\nu}=\left(
\begin{array}{ccc}
 0 & \frac{\chi^2 \zeta \epsilon ^2}{\left(\chi^2+\epsilon ^2\right)} & -\frac{\chi}{2} \\
 \frac{\chi^2 \zeta \epsilon ^2}{\left(\chi^2+\epsilon ^2\right)} & 0 & \frac{\psi}{2} \\
 -\frac{\chi}{2} & \frac{\psi}{2} & \frac{\chi^2 \zeta}{\chi^2+\epsilon ^2} \\
\end{array}
\right) , \nonumber
\ee
\be
K_5^{\mu\nu}=
\left(
\begin{array}{ccc}
 0 & \frac{\chi^2 \zeta^2 \epsilon ^2}{\chi^2+\epsilon ^2} & -\chi \zeta \\
 \frac{\chi^2 \zeta^2 \epsilon ^2}{\chi^2+\epsilon ^2} & 0 & \psi \zeta \\
 -\chi \zeta & \psi \zeta & \frac{\chi^2 \zeta^2}{\chi^2+\epsilon ^2}-\frac{2 \psi}{\chi} \\
\end{array}
\right) .
\ee
The Lagrangian in this coordinates assumes the simple form
\begin{equation}
  L= \frac{\left(\chi(t)^2+\epsilon ^2\right) \left(\chi(t)^2 \zeta'(t)^2-2 \chi'(t) \psi'(t)\right)}{n(t) \chi(t)^2}-n(t) .
\end{equation}
As previously referred, these tensors define further constants of motion as $K_J=K_J^{\mu\nu}p_\mu p_\nu$. Indeed, one can define the corresponding weakly vanishing Hamiltonian
\be
H=N\mathcal{H}=n\left(-\frac{p_{\chi} p_{\psi} \chi^2}{2\left(\chi^2+\epsilon ^2\right)}+\frac{p_{\zeta}^2}{4\left(\chi^2+\epsilon ^2\right)}+1\right)\approx 0
\ee
and explicitly check that the Poisson bracket of the linear charges $Q_I= \xi_I^\mu p_\mu$ as well as the quadratic quantities $K_J$  with $\mathcal{H}$  vanish. The form of all these charges in phase space is
\be
Q_1=p_\zeta, \quad Q_2=p_\psi, \quad Q_3=p_\psi \zeta-\frac{p_\zeta}{\chi}, \qquad Q_h=p_\psi \psi+\frac{p_\zeta\zeta}{2} + \int n(t)dt,
\ee
\be
K_1=\frac{p_\chi^2 \chi^2 \left(\chi^2+\epsilon ^2\right)+2 p_\chi p_\psi \chi \psi \left(\epsilon ^2-\chi^2\right)+\psi \left(p_\psi^2 \psi \left(\chi^2+\epsilon ^2\right)+2 p_\zeta^2 \chi\right)}{2\left(\chi^2+\epsilon ^2\right)},
\ee
\be
K_2=\frac{ p_\chi p_\psi \chi^2 \zeta \left(\chi^2-\epsilon ^2\right)+(p_\chi p_\zeta\chi-p_\psi^2 \chi\psi \zeta+ p_\psi p_\zeta \psi ) \left(\chi^2+\epsilon ^2\right)- p_\zeta^2 \chi^2 \zeta}{\chi \left(\chi^2+\epsilon ^2\right)},
\ee
\be
K_3= \frac{p_\psi \left(p_\chi \chi\left(\epsilon ^2-\chi^2\right)+p_\psi \psi \left(\chi^2+\epsilon ^2\right)\right)+ p_\zeta^2 \chi}{\chi^2+\epsilon ^2},
\ee
\be
K_4=\frac{p_\zeta \left(p_\psi \psi \left(\chi^2+\epsilon ^2\right)+ p_\zeta \chi^2 \zeta\right)-p_\chi \chi \left(p_\zeta \left(\chi^2+\epsilon ^2\right)- 2 p_\psi \chi \zeta \epsilon ^2\right)}{\chi^2+\epsilon ^2},
\ee
\be
K_5= \frac{-2 p_\zeta \chi \zeta \left(\chi^2+\epsilon ^2\right) (p_\chi \chi-p_\psi \psi)+ 2 p_\chi p_\psi \chi^3 \zeta^2 \epsilon ^2 + p_\zeta^2 \left(\chi^3 \zeta^2-2\psi(\chi^2+\epsilon ^2)\right)}{\chi \left(\chi^2+\epsilon ^2\right)} .
\ee
Evaluated on the solution \eqref{solut} the above quantities assume the constant values:
\begin{equation} \label{clasval}
  \begin{split}
    & \mathcal{Q}_1=2 Q\quad \mathcal{Q}_2= 2 c_2 \quad \mathcal{Q}_3=k_3 \\
&   \mathcal{K}_1= \frac{2 \left( m^2+ Q^2+ \epsilon ^2\right)}{c_2^2}, \quad \mathcal{K}_2=\frac{2 k_3 m+4 Q}{c_2}, \quad \mathcal{K}_3= -4 m, \\
    & \mathcal{K}_4=\frac{2 k_3(Q^2+ \epsilon ^2)- 4 m Q}{c_2}, \quad \mathcal{K}_5=\frac{k_3^2 \left(Q^2+ \epsilon ^2\right)-4Q(k_3m + Q)}{c_2^2} .
  \end{split}
\end{equation}

For later use in the quantization procedure we note that the Abelian Poisson subalgebras of dimension three are formed by $(Q_1,Q_2,\mathcal{H})$, $(Q_2,Q_3,\mathcal{H})$, $(Q_1,K_1,\mathcal{H})$, $(Q_1,K_3,\mathcal{H})$, $(K_1,K_4,\mathcal{H})$ and $(K_1,K_5,\mathcal{H})$.
Furthermore, there are Abelian Poisson sub-algebras formed by Killing vector fields and trivial Killing tensors, which are summarized into the following table

\begin{center}
\begin{tabular}{|c|cccccc|}
\hline
& $K_{11}$ & $K_{12}$& $K_{13}$ & $K_{22}$ & $K_{23}$ & $K_{33}$\\
\hline
$Q_1$ & $\surd$ & $\surd$ & $\surd$ &- &-& -\\
$Q_2$ & $\surd$ & $\surd$ & $\surd$ & $\surd$ &$\surd$& $\surd$\\
$Q_3$ & - & - & - & $\surd$ &$\surd$& $\surd$\\
\hline
\end{tabular}
\end{center}
where $K_{IJ}$ are the reducible Killing tensor fields with $K_{IJ}=\frac{1}{2} \left(\xi_I \otimes \xi_J + \xi_I \otimes \xi_I\right)$.

\subsection{Quantization on the pp-wave mini-superspace}

For the quantization, we follow the procedure which is presented in section \ref{section1}. The measure is now $\mu=2\sqrt{2} \frac{\left(\chi^2+\epsilon ^2\right)^{3/2}}{\chi^2}$. Thus, for the hamiltonian constraint we deduce from eq. \eqref{Hamoperator} that

\begin{eqnarray}
& &\widehat{\mathcal{H}}\Psi(\chi,\psi,\zeta)=0\Rightarrow\nonumber\\
& &\hspace{-10mm}\left[-\frac{1}{4 \left( \chi ^2+\epsilon^2\right)}\partial_\zeta^2+\frac{\chi ^3}{ 4\left(\chi ^2+\epsilon^2\right)^2}\partial_\psi+\frac{\chi ^2}{2\left(\chi ^2+\epsilon^2\right)}\partial_\chi\partial_\psi+1\right]\Psi(\chi,\psi,\zeta)=0 , \label{ppWDW}
\end{eqnarray}
while for the first order operators constructed by the Killing vector fields we use the definition \eqref{firstordop}. In order to utilize the irreducible Killing tensor fields in quantization we make use of the so called pseudo-Laplacian operator defined by \eqref{Kop}.

As we mentioned earlier there are six Abelian three-dimensional algebras involving linear and quadratic charges (and of course H). However,  the quantum operators previously defined do not close the corresponding quantum algebras of
$(Q_1,K_1,\mathcal{H})$, $(K_1,K_4,\mathcal{H})$ and $(K_1,K_5,\mathcal{H})$. Thus, we cannot proceed with the quantization of the Abelian sub-algebra involving $K_1$. Thus, we have analyzed the following four cases.

\begin{itemize}

  \item The $(Q_1,Q_2,\mathcal{H})$ algebra: \\
    In the variables $(\chi,\psi,\zeta)$, the vectors $\xi_1$ and $\xi_2$ as we can see from \eqref{kilplushomxyz} are in normal form, thus it is straightforward to derive the common solution of the system of equations
    \begin{equation*}
      \widehat{Q}_1 \Psi (\chi,\psi,\zeta) = \kappa_1 \Psi (\chi,\psi,\zeta), \quad \widehat{Q}_2 \Psi (\chi,\psi,\zeta) = \kappa_2 \Psi (\chi,\psi,\zeta)
    \end{equation*}
    and \eqref{ppWDW}, which reads
    \begin{equation}\label{ppalg1}
   \Psi(\chi,\psi,\zeta) = \frac{C e^{\ima \left(\frac{4 (\chi^2 - \epsilon^2)-\kappa_1^2}{2 \chi \kappa_2}+\kappa_2 \psi +\kappa_1 \zeta\right)}}{(\chi^2+\epsilon ^2)^{1/4}} ,
    \end{equation}
    with $C$ being a constant of integration.

  \item The $(Q_2,Q_3,\mathcal{H})$ algebra: \\
    In this case the supplementary conditions to \eqref{ppWDW} are
    \begin{equation*}
      \widehat{Q}_2 \Psi (\chi,\psi,\zeta) = \kappa_2 \Psi (\chi,\psi,\zeta), \quad \widehat{Q}_3 \Psi (\chi,\psi,\zeta) = \kappa_3 \Psi (\chi,\psi,\zeta)
    \end{equation*}
    and the ensuing solution is
    \begin{equation}\label{ppalg2}
     \Psi(\chi,\psi,\zeta) = \frac{C \sqrt{\chi} e^{ \ima \left(\frac{ \chi \left((\kappa_3-\kappa_2 \zeta)^2+4\right)}{2 \kappa_2}-\frac{2 \epsilon ^2}{\kappa_2 \chi}+\kappa_2 \psi\right)}}{\left(\chi^2+\epsilon ^2\right)^{1/4}} .
    \end{equation}

\item The $(Q_1,K_3,\mathcal{H})$ case: \\
      It is easily proven that $\widehat{K}_3$, as given by \eqref{Kop} commutes with both $\widehat{Q}_1$ and $\widehat{\mathcal{H}}$. Thus we can solve simultaneously the system of equations
      \begin{equation*}
        \widehat{Q}_1\Psi (\chi,\psi,\zeta) = \kappa_1 \Psi (\chi,\psi,\zeta), \quad \widehat{K}_3 \Psi(\chi,\psi,\zeta) = \lambda_1 \Psi (\chi,\psi,\zeta).
      \end{equation*}
      The anticipated solution using the above algebra is
      \begin{equation}\label{solpsi}
        \begin{split}
         \Psi(\chi,\psi,\zeta)=&\frac{\sqrt{\chi} e^{\ima \kappa_1 \zeta}}{\left(\chi^2+\epsilon ^2\right)^{1/4} \sqrt{4(\chi^2- \epsilon ^2)+2 \lambda_1 \chi-\kappa_1^2}}\times\\
      &   \left[C_1 \cos \left(\frac{\sqrt{2 \psi} \sqrt{4(\chi^2-\epsilon ^2)+2 \lambda_1 \chi-\kappa_1^2}}{\sqrt{\chi}}\right)\right. \\
        & \hspace{10mm}\left. +C_2 \sin \left(\frac{\sqrt{2 \psi} \sqrt{4(\chi^2-\epsilon ^2)+2 \lambda_1 \chi-\kappa_1^2}}{\sqrt{\chi}}\right)\right] ,
        \end{split}
      \end{equation}
where $C_1$ and $C_2$ are  integration constants.

In order

  \item The $(Q_2,K_{13},\mathcal{H})$ case: \\
    Here we need to define the operator $\widehat{K}_{13}$ related to the reducible Killing tensor constructed out of $\xi_1$ and $\xi_3$. We just use the two first order operators $\widehat{Q}_1$ and $\widehat{Q}_3$ in the following way
    \begin{equation}
      \widehat{K}_{13} = \frac{1}{2} \left( \widehat{Q}_1 \widehat{Q}_3 + \widehat{Q}_3 \widehat{Q}_1\right)
    \end{equation}
    and a solution to the system $\widehat{Q}_2 \Psi = \kappa_2 \Psi$, $\widehat{K}_{13}\Psi = \kappa_{13} \Psi$ and $\widehat{\mathcal{H}}\Psi =0$ is
      \begin{equation}\label{ppalg3}
      \begin{split}
         \Psi (\chi,\psi,\zeta) = \frac{\chi^{\frac{1}{4}-\frac{\ima\kappa_{13}}{2 \kappa_2}} e^{\frac{2 \ima \left(\chi^2-\epsilon ^2\right)}{ \kappa_2 \chi}+ \ima \kappa_2\psi}}{(\chi^2+\epsilon ^2)^{1/4}} & \left[c_1 H_{-\frac{1}{2}+\frac{\ima \kappa_{13}}{\kappa_2}}\left(\left(\frac{1}{2}+\frac{\ima}{2}\right) \sqrt{\kappa_2} \sqrt{\chi} \zeta\right)+ \right. \\
& \left. c_2 \, _1F_1\left(\frac{1}{4}-\frac{\ima \kappa_{13}}{2\kappa_2};\frac{1}{2};\frac{\ima}{2} \chi \zeta^2 \kappa_2\right)\right].
      \end{split}
    \end{equation}
    The $H_\nu(z)$ and $_1F_1(\alpha;\beta;z)$ are the Hermite function and the Kummer confluent hypergeometric function respectively, while $c_1$ and $c_2$ are integration constants.

\end{itemize}

\subsection{Semiclassical analysis}

In this section we adopt  Bohm's interpretation (see e.g. \cite{struyve,pinto}) of quantum mechanics as a semiclassical approximation for the wave functions found in the preceding section. To this end we have to insert the polar form of the wave function
\be
\Psi(\chi,\psi,\zeta)=\Omega(\chi,\psi,\zeta)e^{iS(\chi,\psi,\zeta)}
\ee
into the Wheeler-DeWitt equation \eqref{ppWDW}, where $\Omega(\chi,\psi,\zeta)$ and $S(\chi,\psi,\zeta)$ are some real functions representing the
magnitude and the phase of the wave function, respectively.

 At this stage, by separating the real and the imaginary part of the equation we end up with the continuity and a modified Hamilton-Jacobi equation respectively. The latter assumes the following form
\be\label{hamjacsem}
\frac{1}{2}G^{\mu\nu}\partial_\mu S\partial_\nu S+1-\frac{\Box \Omega}{2\Omega}=0.
\ee
From this equation we observe that the quantum effects are captured by the last term (called quantum potential), while its vanishing renders this equation classical, in the sense that defining
\be\label{hamjacsem2}
\frac{\partial S}{\partial q^\mu} =p_\mu\equiv\frac{\partial L}{\partial \dot q^\mu}
\ee
gives the classical constraint. Otherwise, when $\Box\Omega$ does not vanish the solutions to the above equations give rise to a geometry inequivalent to the classical solution.

The application of the above procedure to the wavefunctions (\ref{ppalg1},\ref{ppalg2}) yields a vanishing quantum potential and therefore the emerging spacetimes are equivalent to the classical solution. For the third and the fourth case we have to choose the branch of the solution that describes our quantum system and then proceed with the aforementioned analysis.

\subsubsection{Semiclassical analysis of $(Q_1,K_3,\mathcal{H})$ case}

As can be seen from the general wave function \eqref{solpsi}, the form of the quantum solution and thus the resulting semiclassical analysis is highly sensitive to the values of the constants $C_1$ and $C_2$. In principle, in the theory of quantum mechanics, the linear combination of the independent branches of the general solution to the eigenvalue equations -  which actually represents the quantum wave function - is determined by the boundary conditions appropriate to the problem, also guaranteeing the self-adjointness of the operators being used. The single free constant that remains after this process is determined through the normalization of the probability. This is the formal path to take when you have a discrete spectrum problem where the wave function is an element of the Hilbert space.

In cases where the spectrum is continuous the situation is severely more complicated. The wave function does not belong to Hilbert space but rather to its extension, the so called rigged Hilbert space (see \cite{Roberts,Ball} or for an excellent review on the subject \cite{Madrid}). Without getting into many details, the rigged Hilbert space is spanned by the triplet $\Phi \subset H \subset \Psi^\times$, where: $\Phi$ is the subset of the Hilbert space on which the action of the basic operators - up to any order - is well defined (in the sense that the action of the operator on an element of the Hilbert space results in an element of the same space), $H$ is the Hilbert space and $\Phi^\times$ is its dual. The space $\Phi$ is usually spanned by functions that go to zero at the boundary faster than any polynomial. In most known problems of quantum mechanics with continuous spectrum it is some subset of the Schwartz space. On the contrary the elements of the dual space $\Phi^\times$ are allowed to diverge at the boundary as long as the do not diverge faster than any polynomial. This is the space where the wave function of a problem of continuous spectrum ``lives". The generalized spectral theorem guarantees that if an operator $\widehat{Q}$ is self adjoint (over the elements of $\Phi$) then a wave function which is a solution to the eigenvalue problem $\widehat{Q} \Psi = \kappa \Psi$ exists as an element of $\Phi^\times$  \cite{Ball}.

This is the situation which we are facing here. A problem of continuous spectrum where the wave function is not normalizable, but which we do not want to diverge at the border worse than any polynomial. In the particular case of \eqref{solpsi} this does not give us an insight over the constants $C_1$ and $C_2$, but it leads to an interesting analogy between the quantum and the classical solution. If we want a wave function that does not diverge faster that a polynomial in the boundary $\chi\rightarrow \infty$, $\psi\rightarrow \infty$, we need to demand that the argument of the sine and the cosine in \eqref{solpsi} must be always real. In other words we must require
\begin{equation}
  \frac{\psi \left(4(\chi^2-\epsilon ^2)+2 \lambda_1 \chi-\kappa_1^2 \right)}{\chi}\geq 0.
\end{equation}
Let us consider this expression in the original variables $a,b,f$ and at the same time substitute in place of $\lambda_1$ and $\kappa_1$ the classical values that correspond to the conserved charges associated with the quantum operators $\widehat{Q}_1$ and $\widehat{K}_3$ as seen from \eqref{clasval}, i.e. $\kappa_1=2 Q$ and $\lambda_1 = - 4 m$. Then, the previous inequality reads
\begin{equation}
  b^2 \left(a^2-2 a m-Q^2-\epsilon ^2\right) \geq 0.
\end{equation}
The expression in the parenthesis becomes negative when $a$ resides in the region inside the classical horizon
\begin{equation}
  a_{\pm} = m\pm \sqrt{m^2+Q^2+\epsilon ^2} .
\end{equation}
This implies that in order to have a well behaved wave function (for the standards of a continuous spectrum problem) we need to require that in the region $a \in (a_{-},a_{+})$, the variable $b$ is imaginary. But this is exactly the reproduction of what happens in the classical solution. When $a$ takes values inside $(a_{-},a_{+})$ the coefficient of $dz^2$ in the line element changes sign due to $b$ becoming imaginary. In the variables that we are working an imaginary $b$ implies a negative $\psi$, as we can see by the transformation inverse to \eqref{tranppw} which results in $\psi=\frac{a b^2}{2}$. Thus, we can see that this particular classical behaviour that leads to the change of sign in the metric when the effective ``time" variable $a$ crosses the horizon is also reproduced at the quantum level through the demand that the wave function must not diverge badly at the boundary (faster than any polynomial).

Unfortunately, as stated previously, this offers us no hindsight on what to choose over $C_1$ or $C_2$. We can distinguish two major cases that lead to a distinctly different semiclassical behaviour. Firstly, we can see that if we set $C_2=\ima C_1$, then solution \eqref{solpsi} can be written as
\begin{equation}\label{solpsia}
  \begin{split}
  \Psi_1 =&C_1\frac{\sqrt{\chi}}{\left(\chi^2+\epsilon ^2\right)^{1/4} \sqrt{4(\chi^2- \epsilon ^2)+2 \lambda_1 \chi-\kappa_1^2}}\times\\
      & \times   \exp \left[\ima \left(\frac{\sqrt{2 \psi} \sqrt{4(\chi^2-\epsilon ^2)+2 \lambda_1 \chi-\kappa_1^2}}{\sqrt{\chi}}+\kappa_1 \zeta\right) \right].
  \end{split}
\end{equation}
Such a wave function leads to a zero quantum potential, since
\begin{equation}
  \Omega = \frac{\sqrt{\chi}}{\left(\chi^2+\epsilon ^2\right)^{1/4} \sqrt{4(\chi^2- \epsilon ^2)+2 \lambda_1 \chi-\kappa_1^2}} \Rightarrow \Box \Omega = 0.
\end{equation}
As a result, the semiclassical analysis of \eqref{solpsia} does not lead to any emerging quantum correction terms and reproduces exactly the classical space-time.

On the other hand, for any other combination of  $C_1$ and $C_2$, different than the one expressed previously, the only remaining phase of the wave function \eqref{solpsi} is $\kappa_1\zeta$ and the quantum potential is not zero. So, quantum corrections are produced to the trajectories derived with the help of \eqref{hamjacsem2}, which yield
\begin{eqnarray}
-2\frac{\left(\chi^2(t)+\epsilon ^2\right) \psi'(t)}{n(t) \chi^2(t)}=0, \\
-2\frac{\left(\chi^2(t)+\epsilon ^2\right) \chi'(t)}{n(t) \chi^2(t)}=0, \\
\frac{2\left(\chi^2(t)+\epsilon ^2\right) \zeta'(t)- \kappa_1 n(t)}{n(t)}=0.
\end{eqnarray}
The corresponding solution is simply
\be
\chi(t)=\chi_0, \qquad \psi(t)=\psi_0, \qquad \zeta(t)=\frac{\kappa_1 t}{2(\chi_0^2+\epsilon ^2)}+\zeta_0 ,
\ee
where we have gauge fixed $n(t)=1$. This solution creates a  Type III LRS spacetime with  the following line element (after the eliminations of the non-essential constants)

\begin{equation}
\label{statmet}
    ds^2 =  \chi_0^2\left(-dt^2 +dx^2 + e^{-2 x} dy^2 +dz^2\right).
\end{equation}
It is possible to deduce that the space-time is singularity free by using the fact that the Ricci scalar is $R\sim1/\chi_0^2$, the covariant derivative of the Riemmann tensor is zero and finally all fourteen curvature scalars \cite{Thomas} are polynomials of $\frac{1}{\chi_0^2}$.

We observe how different is the situation depending on the choice of constants. On the one hand we have the reproduction of the classical spacetime in  the case where we take $C_2= \ima C_1$ and a singularity free semiclassical space for any other choice. We can take the best from both cases if we consider
\begin{equation}
  \Psi = \begin{cases}
           \Psi_1, &  \chi=a \notin (a_{-},a_{+}) \\
           \Psi_2, &  \chi=a \in (a_{-},a_{+}).
         \end{cases}
\end{equation}
where $\Psi_1$ is given by \eqref{solpsia}, while $\Psi_2$ is the cosine branch of  \eqref{solpsi}
\begin{equation}\label{solpsi2a}
        \begin{split}
         \Psi_2=&\frac{\ima C_1 \sqrt{\chi} e^{\ima \kappa_1 \zeta}}{\left(\chi^2+\epsilon ^2\right)^{1/4} \sqrt{4(\chi^2- \epsilon ^2)+2 \lambda_1 \chi-\kappa_1^2}}\times\\
      &   \cos \left(\frac{\sqrt{2 \psi} \sqrt{4(\chi^2-\epsilon ^2)+2 \lambda_1 \chi-\kappa_1^2}}{\sqrt{\chi}}\right),
        \end{split}
      \end{equation}
where we have choosen $C_2=0$ and appropriately rescaled the other constant of integration, $C_1\rightarrow \ima C_1$, so as to have a $\Psi$ that is continuous at the border formed by the surface $\chi=a_{\pm}$ (something which neither $\Psi_1$ or $\Psi_2$ are solely by themselves). In that way we obtain a wave function that exactly reproduces semiclassically the space-time outside the horizon, while at the same time providing a constant scalar curvature, singularity free, semiclassical description for the inner region, resolving the singularity problem of the classical solution at $\chi=0$.

\subsubsection{Semiclassical analysis of $(Q_2,K_{13},\mathcal{H})$ case}

In this case the corresponding wave function is given by \eqref{ppalg3}. Motivated by the arguments expressed in the previous subsection, we choose that branch of the solution which does not diverge worse than a polynomial at the boundary $\omega= \sqrt{\chi} \zeta \rightarrow \infty$. By the series expansion of the Hermite function we can see that
\begin{equation}
  \begin{split}
    H_{-\frac{1}{2}+\frac{\ima \kappa_{13}}{\kappa_2}}\left(\left(\frac{1}{2}+\frac{\ima}{2}\right) \sqrt{\kappa_2} \omega\right) \approx &   e^{\omega^2} e^{-\ima \frac{\kappa_{13}}{\kappa_2} \ln \omega} \left(\frac{\alpha_1 }{\omega^{1/2}} + \mathcal{O}(\omega^{-3/2}) \right) \\
    & +  e^{\ima \frac{\kappa_{13}}{\kappa_2} \ln \omega} \left(\frac{\alpha_2 }{\omega^{1/2}}  +\mathcal{O}(\omega^{-3/2}) \right) \\
    & \approx \alpha_1 \frac{e^{\omega^2}}{\omega^{1/2}} e^{-\ima \frac{\kappa_{13}}{\kappa_2} \ln \omega}.
    \end{split}
\end{equation}
which means that it diverges badly in terms of an exponential. We thus set $c_1=0$ in \eqref{ppalg3} and consider only the branch of the solution that contains the hypergeometric function. Two distinct approximations are employed: the first concerns  small arguments and, while the second is done for large arguments. The reason for this is to write an approximate expression for \eqref{ppalg3} that allows us to express $\Psi$ in polar form, so that we can use the semiclassical approximation described in the previous section.

\paragraph{Small argument for $a$} which,  if we use the classical solution,
 implies large arguments for the quantity $\omega\equiv\zeta\sqrt{\chi}=\frac{1}{\sqrt{a}}$. The series expansion of the hypergeometric function under consideration yields
\begin{equation} \label{pros1}
  \begin{split}
    {}_1F_1\left(\frac{1}{4}-\frac{\ima \kappa_{13}}{2 \kappa_2};\frac{1}{2};\frac{\ima \kappa_2 }{2} \omega^2\right) \approx \sqrt{\frac{\pi}{\omega}} e^{\frac{\ima \kappa_2}{4} \omega ^2} & \Bigg[ \frac{2^{\frac{1}{4}+\frac{\ima \kappa_{13}}{2 \kappa_2}} (\ima \kappa_2)^{-\frac{1}{4}-\frac{\ima \kappa_{13}}{2 \kappa_2}}}{\Gamma \left(\frac{1}{4}-\frac{\ima \kappa_{13}}{2 \kappa_2}\right)} e^{\frac{1}{4} \ima \kappa_2 \omega ^2} \omega ^{-\frac{\ima \kappa_{13}}{\kappa_2}} \\
    & + \frac{2^{\frac{1}{4}-\frac{\ima \kappa_{13}}{2 \kappa_2}} (-\ima \kappa_2)^{-\frac{1}{4}+\frac{\ima \kappa_{13}}{2 \kappa_2}}}{\Gamma \left(\frac{\ima \kappa_{13}}{2 \kappa_2}+\frac{1}{4}\right)} e^{-\frac{1}{4} \ima \kappa_2 \omega ^2} \omega ^{\frac{\ima \kappa_{13}}{\kappa_2}}
    \Bigg] .
  \end{split}
\end{equation}
Hence, we deduce that the contribution of the hypegeometric function in the phase of \eqref{ppalg3} is just $\frac{\kappa_2\omega^2}{4}$ or equivalently $\frac{\kappa_2 \chi \zeta^2}{4}$, since the expression inside the bracket of \eqref{pros1} is real, as being the sum of two complex conjugate terms.

The full phase of the wave function that is approximate to \eqref{ppalg3}, when $\chi \zeta^2>>1$, is
\begin{equation}
  S_1(\chi, \psi,\zeta) = -\frac{\kappa_{13} }{2 \kappa_2}\ln(\chi) +\frac{2 \chi}{\kappa_2}+\frac{\kappa_2}{4}  \chi \zeta^2-\frac{2 \epsilon ^2}{\kappa_2 \chi}+ \kappa_2 \psi
\end{equation}
and it produces the following equations
\begin{align} \label{seq1}
n(t) \left(-2 \kappa_{13} \chi(t)+\chi(t)^2 \left(\kappa_2^2 \zeta(t)^2+8\right)+8 \epsilon ^2\right)+8 \kappa_2 \left(\chi(t)^2+\epsilon ^2\right) \phi'(t)=0, &\\ \label{seq2}
-\kappa_2-\frac{2 \left(\chi(t)^2+\epsilon ^2\right) \chi'(t)}{n(t) \chi(t)^2}=0 , &\\ \label{seq3}
\frac{2 \left(\chi(t)^2+\epsilon ^2\right) \zeta'(t)}{n(t)}-\frac{\kappa_2}{2}  \chi(t) \zeta(t)=0 .&
\end{align}
The resulting line element  is given by
\begin{align} \label{semet1}
  ds^2 = - & \left(1 +\frac{\psi_0\kappa_2^2}{2a}-\frac{\epsilon^2}{a^2}+ \frac{\psi_1 \kappa_2^2}{2} \frac{\ln(a)}{a} \right)^{-1} da^2  \nonumber \\
 &+ a^2 \left(dx^2 + e^{-2x} dy^2\right) + \frac{4}{\kappa_2^2} \left(1 +\frac{\psi_0\kappa_2^2}{2a}-\frac{\epsilon^2}{a^2}+ \frac{\psi_1 \kappa_2^2}{2} \frac{\ln(a)}{a} \right) dz^2 ,
\end{align}
where $\psi_0$ and $\psi_1$ are integration constants. This space is not singularity free because it has non-constant scalar curvatures.

\paragraph{Large arguments a.} On the other hand if we consider large arguments for scale factor we have to use small arguments for $\omega$. The approximate form of wave function is
\be
\Psi(\chi,\psi,\zeta)\approx \left(\frac{\chi}{\chi^2+\epsilon ^2} \right)^{\frac{1}{4}}e^{\ima\left[ \kappa_2 \psi+\frac{2\left(\chi^2-\epsilon ^2\right)}{ \kappa_2 \chi}-\frac{ \kappa_{13} \ln (\chi)}{2\kappa_2}\right]}
\ee
The corresponding semiclassical equations are
\begin{align}
-\frac{2\left(\chi^2(t)+\epsilon ^2\right) \psi'(t)}{n(t) }-\frac{4 \left( \chi(t)^2+\epsilon ^2\right)-\kappa_{13} \chi(t)}{ 2\kappa_{2} }= & 0 ,\\
- \kappa_{2}-\frac{2\left(\chi^2(t)+\epsilon ^2\right) \chi'(t)}{n(t) \chi^2(t)}= &0 , \\
\frac{2\left(\chi^2(t)+\epsilon ^2\right) \zeta'(t)}{n(t)}=&0 .
\end{align}
By integrating the equations, the ensuing line element becomes
\begin{align}
ds^2=-\left(1+\frac{\psi_0\kappa_2^2}{2a}- \frac{\epsilon ^2}{a^2}-\frac{\kappa_{13}}{4} \frac{  \ln (a)}{a}\right)^{-1}da^2+a^2 \left(dx^2 + e^{-2x} dy^2\right)+ &\nonumber\\
\frac{4}{\kappa_2^2}\left(1+\frac{\psi_0\kappa_2^2}{2a}- \frac{\epsilon ^2}{a^2}-\frac{\kappa_{13}}{4} \frac{  \ln (a)}{a}\right)dz^2,  &\label{linelemfour2}
\end{align}
where again $\psi_0$ is an integration constant. This is similar to the line element \eqref{semet1} from the previous case. The corresponding space is not singularity free but the region under consideration is far away from singularity, so it is not necessary to be singularity free.

At this point we have to note that in both semiclassical metrics \eqref{semet1} and \eqref{linelemfour2} we have kept all terms in $a$ irrespectively of the approximation we considered for the arguments of the hypergeometric function. The approximations $\zeta\sqrt{\chi}>>1$ and $\zeta\sqrt{\chi}<<1$ are being adopted so that we can approximate the polar form of the wave function. The relation that $\omega=\zeta\sqrt{\chi}$ has with the classical $a$, $\omega=\frac{1}{\sqrt{a}}$, that we use as an effective time variable in the classical solution is valid only on mass shell. In that respect, and although an absolute comparison between the semiclassical and the classical metrics is not possible due to the fact that they are distinct geometries, we can comment on how the dependence of the coefficients of the metric changes with respect to the ``time" variable $a$. However, we have to keep in mind that in each case we are talking for two different intrinsic times $a_{cl}$ and $a_{sc}$ for the classical and semiclassical manifolds respectively. We can see that both of the semiclassical metrics yield a term proportional to $\frac{\ln(a)}{a}$ in place of the $\frac{Q^2}{a^2}$ appearing in the classical metric. So, we see that the main contribution of the quantum corrections has to do with the effect of the contribution of the electric field inside the line element.

It is also interesting to make the association of the quantum numbers $\kappa_2$ and $\kappa_{13}$ with their classical counterparts that we can deduce from \eqref{clasval}. The constant $\kappa_2$ is associated with $Q_2$ and consequently with the classical constant of integration $k_2$ which is a non-zero, non-essential constant for the geometry. The constant $\kappa_{13}$ corresponds to the classical conserved quantity constructed by the trivial Killing tensor $K_{13}= \frac{1}{2}\left(\xi_1 \otimes \xi_3 + \xi_3 \otimes \xi_1 \right)$, which on mass shell assumes the value $2 Q k_3$ (again by relations \eqref{clasval}). Notice however that $k_3$ is also a non-essential constant which defines the value of the potential at infinity as seen by \eqref{solut}. If we make this association with the classical level and consider $k_3=0\Rightarrow \kappa_{13}=2 Q k_3=0$, then the second semi-classical metric \eqref{linelemfour2} - for large $a$ - has no contribution from the electric field, while on the other hand the semi-classical space-time \eqref{semet1} that describes space-time in small $a$ always has a contribution of the form $\frac{\psi_1 \kappa_2^2}{2} \frac{\ln(a)}{a}$ combined with a semiclassical electric potential which in this case is (out of solving the system \eqref{seq1}-\eqref{seq3})
\begin{equation}
  f = \sqrt{\frac{2}{a}\left(\frac{\kappa_{13}}{\kappa_2^2}+2 \psi_1\right)}.
\end{equation}
This means that you always get a contribution from $f\neq 0$ inside the metric, when $a<<1$ even if $\kappa_{13}=0$. In contrast to what happens for $a>>1$, where the condition $\kappa_{13}=0$ can extinguish the contribution from the metric. As far as the magnetic charge is concerned we see that its contribution is unaltered though this description. The latter may have its origin in the fact that the magnetic charge appears fixed inside the Hamiltonian function and it is not treated as a product of a direct degree of freedom like the electric charge stemming from $f$ in $A_\mu$.

\section{The case of a single magnetic monopole}

If we wish to consider the case where the electric potential $f(t)$ is zero, then the process has to be resumed from the Lagrangian level. This is because the mini-superspace geometry is different: the number of configuration degrees of freedom reduces to two i.e. $(a,b)$ and in the constant potential parametrization the mini-supermetric reads
\begin{equation} \label{minimagn}
  G_{\alpha\beta} = -\frac{2b(a^2+\epsilon^2)}{a}\begin{pmatrix}
                      \frac{b}{a} & 1 \\
                      1 & 0
                    \end{pmatrix}
\end{equation}
and it is that of a flat space. The corresponding Lagrangian is given by \eqref{lagconpot}. With the help of the integrals of motion constructed by the three Killing vector fields
\begin{equation} \label{kilmagn}
  \xi_1 = \frac{a \left(a^2-\epsilon^2\right)}{a^2+\epsilon^2} \partial_a -\frac{a^2 b}{a^2+\epsilon^2} \partial_b, \quad \xi_2 = -\frac{a^2}{2(a^2+\epsilon^2)} \partial_a +\frac{a b}{ 4(a^2+ \epsilon^2)} \partial_b, \quad \xi_3 = \frac{1}{ a b}\partial_b
\end{equation}
and the homothetic vector (that provides a non-local conserved charge)
\begin{equation}
  \xi_h = \frac{b}{2} \partial_b
\end{equation}
the classical solution can be derived easily. The line element is the same as the solution \eqref{linelcl} with $Q=0$. Of course $m$ can still be considered either positive or negative and an horizon is formed at
\begin{equation}
  a_{\pm} = m \pm \sqrt{m^2+\epsilon^2}
\end{equation}
depending on whether we are in the $a\in \mathbb{R}_{+}$ region or in the $a\in \mathbb{R}_{-}$ respectively.

In what regards the on mass shell values of the three conserved charges corresponding to \eqref{kilmagn}, we can see that only $Q_1$ is associated with an essential constant of the geometry yielding $Q_1 =2 m$, while $Q_2=-\frac{1}{2}$ and $Q_3=2$. To be more precise, $Q_2$ and $Q_3$ are associated with non-essential constants of the geometry that are bound to satisfy $Q_2 Q_3=-1$ due to the constraint. The particular values that we give here, are those that emerge from the line-element from which the non-essential constants are cleared out.

\subsection{Quantization}

In order to proceed with the quantization it is useful to adopt the transformation $(a,b)\rightarrow (\chi,\psi)$ with\footnote{We understand that here the variables $\chi$, $\psi$ and $\zeta$ are different from the ones used in the previous section.}
\begin{equation}
  a= \frac{1}{4} \left(-\chi+\sqrt{16 \epsilon^2+\chi^2}\right), \quad b = \frac{2\sqrt{2 \psi}}{\left(\sqrt{16 \epsilon^2+\chi^2}-\chi\right)^{\frac{1}{2}}}
\end{equation}
that makes the mini-superspace metric reads $G_{\alpha\beta} = \begin{pmatrix}0 & 1\\ 1 & 0\end{pmatrix}$. In these coordinates it can be seen that the vectors \eqref{kilmagn} are transformed into
\begin{equation} \label{kilmagn2}
  \xi_1 = \chi \partial_\chi - \psi \partial_\psi, \quad \xi_2 = \partial_\chi, \quad \xi_3 =\partial_\psi .
\end{equation}

The corresponding operators give rise to two Abelian subalgebras involving $(\widehat{Q}_2,\widehat{Q}_3,\widehat{\mathcal{H}})$ and $(\widehat{Q}_1,\widehat{\mathcal{H}})$. The first case, as can be seen by the form of \eqref{kilmagn2}, leads to a plain wave solution for the wave function
\begin{equation}
  \psi = C e^{\ima (\kappa_2 \chi+\kappa_3 \psi)}
\end{equation}
with the Wheeler-DeWitt equation, $\widehat{\mathcal{H}}\Psi=0$, implying the eigenvalue constraint $\kappa_2 \kappa_3 = -1$.

The second case proves to be more interesting; the solution of $\hat Q_1 \Psi = \kappa_1 \Psi$ implies that
\begin{equation}
  \Psi(x,y) = \chi^{\ima \kappa_1} \psi (\chi \psi).
\end{equation}
The imposition of the Hamiltonian constraint  $\widehat{\mathcal{H}}\Psi=0$ fixes the function $\psi (x y)$ to be
\begin{equation}
   \psi (\chi \psi) = (\chi \psi)^{-\frac{\ima \kappa_1}{2}} \left(c_1 I_{\ima \kappa_1}\left(2 \sqrt{\chi \psi}\right)+c_2 I_{-\ima \kappa_1}\left(2 \sqrt{\chi \psi}\right)\right).
\end{equation}
Hence, the full wave function becomes
\begin{equation} \label{wavemagn}
  \Psi (\chi, \psi) = \left(\frac{x}{y}\right)^{\frac{\ima \kappa_1}{2}} \left(c_1 I_{\ima \kappa_1}\left(2 \sqrt{\chi \psi}\right)+c_2 I_{-\ima \kappa_1}\left(2 \sqrt{\chi \psi}\right)\right),
\end{equation}
where $I_{\nu}(\zeta)$ is the modified Bessel function of the first kind. By going back to our initial variables $(a,b)$ we can see that
\begin{equation} \label{argumentmagn}
  \zeta^2=\chi \psi = b^2 \left(\epsilon^2-a^2\right),
\end{equation}
which means that we have to be careful on whether the product $\zeta^2=\chi \psi$ is positive or negative since it changes significantly the behaviour of $I_{\nu}(\zeta)$. For example: if $\zeta\rightarrow \infty$ the $I_{\nu}(\zeta)$ diverges, but when $\zeta\rightarrow \ima \infty$, $I_{\nu}(\zeta)\rightarrow 0$. From \eqref{wavemagn} we can also observe that we obtain two branches, which however can be interchanged by a catoptric transformation $\kappa_1\rightarrow -\kappa_1$. Thus, we need only consider from now on only the first branch. In what follows we utilize once more Bohm's interpretation of quantum mechanics as a tool to provide a semi-classical approximation by introducing the notion of a trajectory at this level.

\subsection{Semi-classical approximation}

In what follows we consider the contribution of the first part of \eqref{wavemagn} in the wave function, i.e. we take $c_2=0$. Let us start by studying the behaviour of the argument $\zeta$ in \eqref{argumentmagn}, which depends on the behaviour  of $a$ and $b$. If we keep $b$ fixed, it can be easily seen that for $a\rightarrow \pm \infty$, $\zeta^2$ approaches $-\infty$ and as a result we have $\zeta$ being imaginary. The approximate expression for the particular Bessel function when the argument is purely imaginary $\zeta=\ima s$ yields
\begin{equation}
  I_{\ima \kappa_1}(\ima s) \approx \frac{\cos \left[ s -\left(\frac{\pi }{4}+\frac{\ima}{2}  \kappa_1 \pi  \right)\right]}{s^{1/2}}+ \mathcal{O}(s^{-3/2}),
\end{equation}
which means that the Bessel function does not contribute to the phase and we may approximate the wave function \eqref{wavemagn} with
\begin{equation}\label{magnmonowf}
  \Psi (a,b) \approx \frac{1}{b \sqrt{a^2-\epsilon^2}} \exp \left[ \frac{\ima\kappa_1}{2} \ln \left(\frac{4 \left(a^2-\epsilon^2\right)}{a^2 b^2}\right) \right] .
\end{equation}
By having brought the approximate wave function in polar form we can use the formulas \eqref{hamjacsem2} to extract the semiclassical trajectories. The latter are governed by equations:
\begin{subequations}  \label{semieqmagn}
  \begin{align}
    \frac{\partial L}{\partial a'} & = \frac{\kappa_1 \epsilon ^2}{a(t)(a(t)^2-\epsilon^2) } ,\\
    \frac{\partial L}{\partial b'} & = -\frac{\kappa_1}{ b},
  \end{align}
\end{subequations}
where $L$ we have to use is obtained through \eqref{generalLag2} by setting \eqref{minimagn} as the mini-superspace metric.

At this point we have two equations and three functions to be determined, $a(t)$, $b(t)$ and $n(t)$. This set of equations can be easily integrated to yield a semi-classical line element of the form
\begin{equation} \label{sem1onlym}
  ds^2 = - \frac{4 b_0^2 a^2 }{\kappa_1 (a^2-\epsilon^2)} da^2 + a^2 (dx^2 + e^{-2x} dy^2) + \frac{b_0^2}{a^2 - \epsilon^2} dz^2 .
\end{equation}
The resulting spacetime exhibits curvature singularities at $a=0$ and $a= \pm \epsilon$. It is interesting to note that the exact same result can be reached by considering the region $a\rightarrow 0$, $b\rightarrow \infty$, which implies that $\zeta \in \mathbb{R}$ with $\zeta\rightarrow \infty$. Then, the approximate wave function still does not have any contribution in the phase rising from the Bessel function, since
\begin{equation}
  I_{\ima \kappa_1} (\zeta) \approx \frac{1}{\zeta} \left(e^{\zeta} + e^{-\zeta- \kappa_1\pi } \right)
\end{equation}
and, as a result, one arrives at the same semiclassical line element \eqref{sem1onlym}.

What we have checked until now is the semiclassical behaviour as $\zeta^2$ assumes great values. It is interesting to see what happens in the other region when $\zeta^2 \rightarrow 0$. By considering a finite $b$, we deduce from \eqref{argumentmagn}, that $a\rightarrow 0$ means that $\zeta^2$ is small but positive, thus $\zeta\in \mathbb{R}$. The modified Bessel function $I_\nu (\zeta)$ in this case can be written in terms of the series
\begin{equation}
  I_\nu (\zeta) = \left(\frac{\zeta}{2}\right)^{\nu} \sum_{k=0}^{+\infty} \frac{(\zeta^2/4)^{2k}}{k!(\Gamma(\nu+k+1))},
\end{equation}
where $\Gamma$ stands for the Gamma function. Since $\zeta$ is now real we can immediately see that the only contribution in the phase, which plays a role in the derivation of the semiclassical trajectories, is given by $\zeta^{\nu}=\zeta^{\ima \kappa_1}$. As a result the wave function, if we consider that $c_2=0$, can approximated by
\begin{equation}
  \Psi (a,b) \approx \Omega (a,b) e^{\ima \kappa_1 \ln \left(\frac{2(\epsilon ^2-a^2)}{a}\right)}
\end{equation}
and the corresponding semiclassical equations are
\begin{subequations}
  \begin{align}
    -\frac{2\left(a(t)^2 +\epsilon^2\right) b(t)}{a(t) n(t)} \left(\frac{b(t)a'(t)}{a(t)} +b'(t)\right) + \frac{\kappa_1 \left(a(t)^2+\epsilon ^2\right)}{a(t)(\epsilon ^2 -a(t)^2)}&=0, \\
    -\frac{2 b(t) \left(a(t)^2+\epsilon ^2\right) a'(t)}{a(t) n(t)}& =0 .
  \end{align}
\end{subequations}
This system can be solved easily and in terms of $a(t)$ and $n(t)$ and the solution reads
\begin{equation} \label{semsolmagn1}
a(t)=a_0, \quad  n(t) =-\frac{2 \left(a_0^2-\epsilon ^2\right) b(t) b'(t)}{\kappa_1} .
\end{equation}
Finally, the corresponding semiclassical line element (after the elimination of the non-essential constants) is
\begin{equation}
  ds^2 = \alpha_{0}^2\left(-db^2+dx^2+e^{-2x}dy^2+b^2dz^2\right) .
\end{equation}
All the scalar curvature of this space are constant, thus it is singularity free. As also happened in the large $\zeta$ section the same result can be reached by considering $a>\epsilon$, but $b\rightarrow 0$, which results in a small but purely imaginary $\zeta=\ima s$. The approximation over the Bessel function for $s<<1$ yields $I_{\ima \kappa_1}(\ima s) \sim s^{\ima \kappa_1}$ producing the exact same phase correction.

\section{A single electric charge}

As happens in the case of the magnetic monopole, a distinct - from the general case - mini-superspace also emerges by considering a single electric charge. Indeed, if we set $\epsilon=0$ in the general Lagrangian \eqref{lagconpot}, we get
\begin{equation} \label{Lagelecric}
  L = -\frac{1}{n}\left(2 a b a' b' +b^2 a'^2-a^2 f'^2\right)- n
\end{equation}
from which we derive the following mini-superspace metric
\begin{equation} \label{miniflat}
  G_{\mu\nu} = -2 \begin{pmatrix}
                 b^2 &  a b & 0 \\
                  a b & 0 & 0 \\
                 0 & 0 & -a^2
               \end{pmatrix} .
\end{equation}
This time the mini-superspace is three dimensional, since we keep $f$ as a degree of freedom, but in contrast to the $\epsilon\neq 0$ case, it is flat.

At this point a comment is in order. The quantization of a mini-superspace with a metric similar to \eqref{miniflat} has been considered in \cite{dimet2017}. It is interesting to note, that in general the LRS Bianchi type III possesses a familiar mini-superspace structure to that of a static spherically symmetric line element. The difference of course being that the former has as an independent variable the time $t$, while the latter evolves with respect to the radial distance instead. An additional difference is in the sign of the kinetic term owed to the electric field $f$ in the Lagrangian. It can be seen in (3.5) of \cite{dimet2017} that the corresponding Lagrangian (if you discard an overall minus) has a ``$-\frac{1}{n}a^2 \dot{f}^2$" term instead of the ``$+\frac{1}{n}a^2 \dot{f}^2$" appearing in \eqref{Lagelecric}. However, this does not affect to a great extent the process up to the point of the derivation of the Wheeler-DeWitt equation and this is why we are only going to sketch the basic points, following exactly the spirit of \cite{dimet2017}.

Since the Leorentzian three dimensional mini-superspace appearing here is flat, we know that there exist six first order integrals of motion consisting of three translations, two pseudorotations and one actual rotation. The last three being the elements of the well known $SO(2,1)$ group. An interesting algebra to use in the quantization procedure is the one that is formed by the true rotation, say $Q_6$, and the Casimir invariant $Q_{Cas}$ of the $\mathfrak{so}(2,1)$ algebra. By performing the transformation
\begin{equation}
  \begin{split}
   a &= \frac{1}{4} u (\sinh (\vn) \cos (w)+\cosh (\vn)),  \quad b= \frac{2 \sqrt{2}}{\sinh (\vn) \cos (w)+\cosh (\vn)},\\
   f & = \frac{2 \sqrt{2} \sin (w)}{\coth (\vn)+\cos (w)}
  \end{split}
\end{equation}
the operator corresponding to the rotation $Q_6$ is being brought into normal form in the $w$ coordinate. The common solution of the system of eigenequations
\begin{equation}
  \widehat{Q}_6 \Psi = k \Psi, \quad \widehat{Q}_{Cas}\Psi = \ell (\ell+1) \Psi, \quad \mathcal{H}\Psi =0
\end{equation}
is
\begin{equation}\label{psielectric}
  \Psi =e^{\ima k w}\left(C_1 P_{\ell}^k (\cosh(\vn))+C_2 Q_{\ell}^k (\cosh(\vn))\right) \left(C_3 j_\ell(\sqrt{2} u)+C_4 \mathrm{y}_\ell(\sqrt{2} u)\right),
\end{equation}
where $P_{\ell}^k$, $Q_{\ell}^k$, $j_\ell$ and $\mathrm{y}_\ell$ are the associated Legendre and spherical Bessel functions of the first and second kind respectively.

Due to $\widehat{Q}_6 = -\ima \frac{\partial}{\partial w}$ corresponding to a true rotation in this flat mini-superspace, we can impose a periodical boundary condition in the $w$ variable that leads to a discrete spectrum for $k$, i.e. $k\in \mathbb{Z}$. As also happens for the study of the Reissner - Nordstr\"om geometry in \cite{dimet2017}, $Q_6$ is classically  related to the electric charge, while $Q_{Cas}$ with a quadratic combination of the charge $Q$ and the ``mass" $m$. In particular,
upon the classical solution
\begin{equation}
  ds^2 = \left(1-\frac{2m}{a}-\frac{Q^2}{a^2}\right)^{-1} da + a^2 \left(dx^2+e^{-2x} dy^2\right) +\left(1-\frac{2m}{a}-\frac{Q^2}{a^2}\right) dz^2
\end{equation}
and $f(t) = \frac{Q}{a}$, we can see that
\begin{equation}
  Q_6 \propto Q, \quad Q_{Cas} \propto - (m^2+Q^2)
\end{equation}
from which we observe that $Q_{Cas}<0$. This latter condition is not necessarily true for the Reissner - Nordstr\"om case in \cite{dimet2017} where $Q_{Cas}$ could also assume positive values. The negativity of the Casimir invariant implies that at a quantum level we should also expect $\ell(\ell+1)<0$. This can happen by assuming a complex quantum number $\ell=-\frac{1}{2}+\ima s$, $s\in \mathbb{R}$.

\section{Discussion}

In the present work we have studied the mini-superspace model of a Bianchi type III LRS geometry coupled to a source-free electromagnetic field.

 At the classical level we have found the solution space by using the symmetries of the configuration manifold.
Specifically, our first step was to find the most general electromagnetic tensor compatible with the symmetries of the Bianchi type III LRS line element. To this end we have considered the algebraic restrictions imposed on $T_{\mu\nu}$ through Einstein's equations when the geometry is as above. The result, when we also use the Maxwell equations, is an electromagnetic field tensor with only $E_3(t)$ and $B_3(x)=\epsilon \, e^{-x}$. The final solution is given by \eqref{lineemel}. The mini-superspace  description of the model is reached by applying the four Killing fields \eqref{killfield} to a general $F_{\mu\nu}(t,x,y,z)$ and demanding that it is expressed in terms of an electromagnetic potential. A noteworthy fact is that the valid Lagrangian obtained defines a mini-superspace  which is a 3D pp-wave for the general case admitting five non-trivial Killing tensors and three Killing vector fields. The cases of electric and/or magnetic field only acquire flat mini-superspaces of dimension three and two respectively.

At the quantum level we have followed the canonical  quantization method. For determining a unique wave function we have used, except the Wheeler-DeWitt constraint equation, the operators corresponding to Abelian subalgebras formed by the various  linear and quadratic symmetry generators.
In each of the three cases of mini-superspace  a semi-classical analysis of the solutions is done using Bohm' s interpretation of quantum mechanics. A noticeable  fact is that some of the solutions  give singularity free semiclassical space-times.

At this point, we have to notice that in the procedure of semiclassical analysis the continuity equation is not always valid.
This fact emerges due to the approximations needed to bring the wave functions in polar form.
This problem is overcome if we apply the corresponding limit also to the continuity equation. For example, this happens in eq \eqref{magnmonowf}, with the corresponding continuity equation
\be
g^{\mu\nu}\nabla_\mu S\nabla_\nu \Omega+\frac{\Omega}{2}\Box S=
\frac{4 \epsilon ^2+\kappa_1 \epsilon ^2}{b^2 \left(a^4-\epsilon ^4\right)}+\frac{4 a^2}{b^2 \left(a^4-\epsilon ^4\right)},
\ee
which is going to zero if we consider that orders of the form $\frac{1}{a^n}$ with $n>1$ are negligible.

One step beyond  our procedure could be a modification of the quantization method. Specifically, the use of pseudo-Laplacian to define the operators corresponding to the quadratic charges is not the most general thing one can do.
The most general operator, which satisfies the criteria of being scalar, second order, natural and hermitian, is given by
\begin{align*}
\hat K \Psi=(1-a)\left[\phi_1K^{ab}\nabla_a\left(\frac{1}{\phi_1\phi_2}\nabla_b(\phi_2\Psi)\right)+\phi_2\nabla_a\left(\frac{1}{\phi_1\phi_2}\nabla_b(\phi_1K^{ab}\Psi)\right)\right] & \\
+a\nabla_a\left(K^{ab}\nabla_b(\phi_1\Psi)\right), &
\end{align*}
where $\phi_1$ and $\phi_2$ are scalars constructed out of the metric and the Killing tensor $K_{ab}$. Even the closure of the quantum algebras can be improved, thus allowing for more possibilities \cite{winp}.

\acknowledgments{
N. Dimakis acknowledges financial support by FONDECYT postdoctoral grant no. 3150016. The co-author T. Pailas thanks the General Secretariat for Research and Technology (GSRT) and the Hellenic Foundation for Research and Innovation (HFRI) of the Greek Ministry of Education for supporting his PhD fellowship.}

\end{document}